\documentclass[aip,amsmath,amssymb,reprint,]{revtex4-2}

\usepackage{graphicx}
\usepackage{dcolumn}
\usepackage{bm}

\usepackage[utf8]{inputenc}
\usepackage[T1]{fontenc}
\usepackage{mathptmx}
\usepackage{etoolbox}
\usepackage{xcolor}

\makeatletter
\def\@email#1#2{%
	\endgroup
	\patchcmd{\titleblock@produce}
	{\frontmatter@RRAPformat}
	{\frontmatter@RRAPformat{\produce@RRAP{*#1\href{mailto:#2}{#2}}}\frontmatter@RRAPformat}
	{}{}
}%
\makeatother
\begin{document}
	
	\preprint{AIP/123-QED}
	
	\title{Mechanical mode engineering with orthotropic metamaterial membranes}
	
	\author{G. Conte}
	\affiliation{Dipartimento di Fisica "E. Fermi", Università di Pisa, Largo B. Pontecorvo 3, 56127 Pisa - Italy}
	\affiliation{NEST, CNR - Istituto Nanoscienze and Scuola Normale Superiore, Piazza San Silvestro 12, 56127 Pisa - Italy}
	\author{L. Vicarelli}
	\email{leonardo.vicarelli@nano.cnr.it}
	\affiliation{NEST, CNR - Istituto Nanoscienze and Scuola Normale Superiore, Piazza San Silvestro 12, 56127 Pisa - Italy}
		\author{S. Zanotto}
	\affiliation{NEST, CNR - Istituto Nanoscienze and Scuola Normale Superiore, Piazza San Silvestro 12, 56127 Pisa - Italy}
	\author{A. Pitanti}
	\affiliation{NEST, CNR - Istituto Nanoscienze and Scuola Normale Superiore, Piazza San Silvestro 12, 56127 Pisa - Italy}
	
	\date{\today}
	
\begin{abstract}
Metamaterials are structures engineered at a small scale with respect to the wavelength of the excitations they interact with. These structures behave as artificial materials whose properties can be chosen by design, mocking and even outperforming natural materials and making them the quintessential tool for manipulation of wave systems. In this Letter we show how the acoustic properties of a silicon nitride membrane can be affected by nanopatterning. The degree of asymmetry in the pattern geometry induces an artificial anisotropic elasticity, resulting in the splitting of otherwise degenerate mechanical modes. The artificial material we introduce has a maximum Ledbetter-Migliori anisotropy of 1.568, favorably comparing to most bulk natural crystals. With an additional freedom in defining arbitrary asymmetry axes by pattern rotation, our approach can be useful for fundamental investigation of material properties as well as for devising improved sensors of light, mass or acceleration based on micromechanical resonators. 
\end{abstract}
	
\maketitle
	
Modern strategies for the manipulation of electronic, photonic or acoustic waves rely on the concept of creating artificial materials by deterministic patterning. The fine structure of geometrical elements with a size smaller than the system wavelength cannot be resolved, yet it can manifest in macroscopic effects resulting in the creation of effective artificial materials with properties chosen by design. Recently, the concept of artificial material has entered into the mechanical realm, focusing on frequencies around and exceeding the audible spectrum; both acoustic metamaterials\cite{acoustic_metamaterials_review} and metasurfaces\cite{acoustic_metasurfaces_review} have been introduced, with size and periodicity of scattering elements significantly smaller than the acoustic wavelength. Artificial acoustic materials have revealed interesting properties such as negative refraction\cite{negative_refraction}, superlensing\cite{acoustic_superlens} and cloaking\cite{acoustic_cloak} of mechanical waves.

Along with micro- and nano-structuration, the miniaturization of whole mechanical resonators has been pushed to the micrometric scale; among the others, a wide investigation has interested nanometer-thick, micrometer-wide membranes\cite{nanomembrane1, nanomembrane2, nanomembrane3}. Thanks to their large quality factors\cite{nanomembrane1, Koppens, spiderweb} and extreme aspect ratio, these kinds of device have been used as a standard platform for classical and quantum effects in optomechanics\cite{Koppens,Thompson} and for light\cite{Vicarelli}, pressure\cite{Dantan}, mass\cite{mass} and other sensing applications\cite{nanomembrane2, Lemme}. The mechanical membranes are an ideal platform for hosting photonic metasurfaces, which can be introduced by periodically patterning a portion of the device. This merging has enabled the realization of a class of optomechanical devices where membrane mechanical modes affect the electromagnetic modes in the photonic structure \cite{Kippenberg}. Such devices have been very successful in photonics, where the mechanical actions added a dynamic component to the optical response of static devices\cite{Robert-Philip, Krenner}; applications included high-performance reflectors\cite{Lawall, Deleglise, Groblacher}, metamaterial absorbers\cite{Cubukcu} and manipulation of light polarization\cite{Zanotto}, to name just a few. Similar concepts can be translated to mechanical membranes with embedded acoustic metasurfaces, where the mechanical resonator can be tweaked by controlling the properties of its constitutive (artificial) material.

In this Letter we report on square silicon nitride (Si$_3$N$_4$) membranes periodically patterned with holes which defined phononic metasurfaces. We realized devices with a different degree of asymmetry in the pattern geometry, resulting in degeneracy breaking for selected mechanical modes, which we characterized through self-mixing interferometry. The asymmetric pattern made the metasurface equivalent to an artificial homogeneous orthotropic shell where the axis anisotropy can be controlled with the geometrical asymmetry. Furthermore, our design allowed to arbitrarily define the orthotropic axes which could be oriented in any direction on the membrane plane, obtaining devices whose realization would be challenging employing natural anisotropic materials. 
\begin{figure}[h]
	\includegraphics[width=8cm]{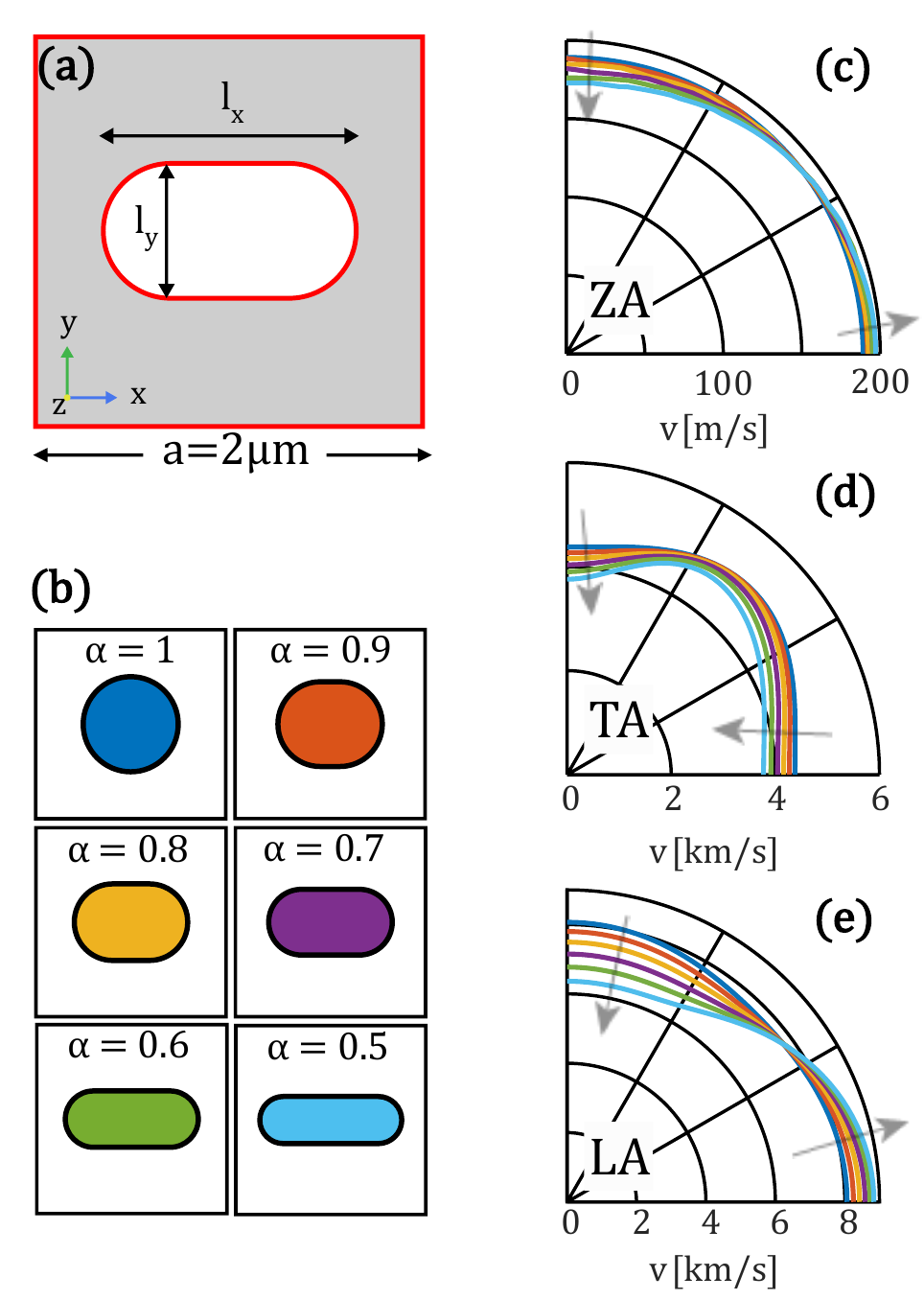}
	\caption{(a) Unit cell of the metasurface. The central hole shape can be controlled by the asymmetry parameter $\alpha$, going from isotropic ($\alpha=1$) to a maximum anisotropy before structural fail ($\alpha=0.5$), panel (b). (c-e) Polar plots of sound velocities at 2 MHz of the first three acoustic modes for different $\alpha$s. The full angle velocity plots can be obtained by mirror symmetry at 0$^{\circ}$ and 90$^{\circ}$. The different lines correspond to the unit cells of panel (b) of the same color.}
    \label{fig:1}
\end{figure}
\begin{figure*}[t]
	\includegraphics[width=17cm]{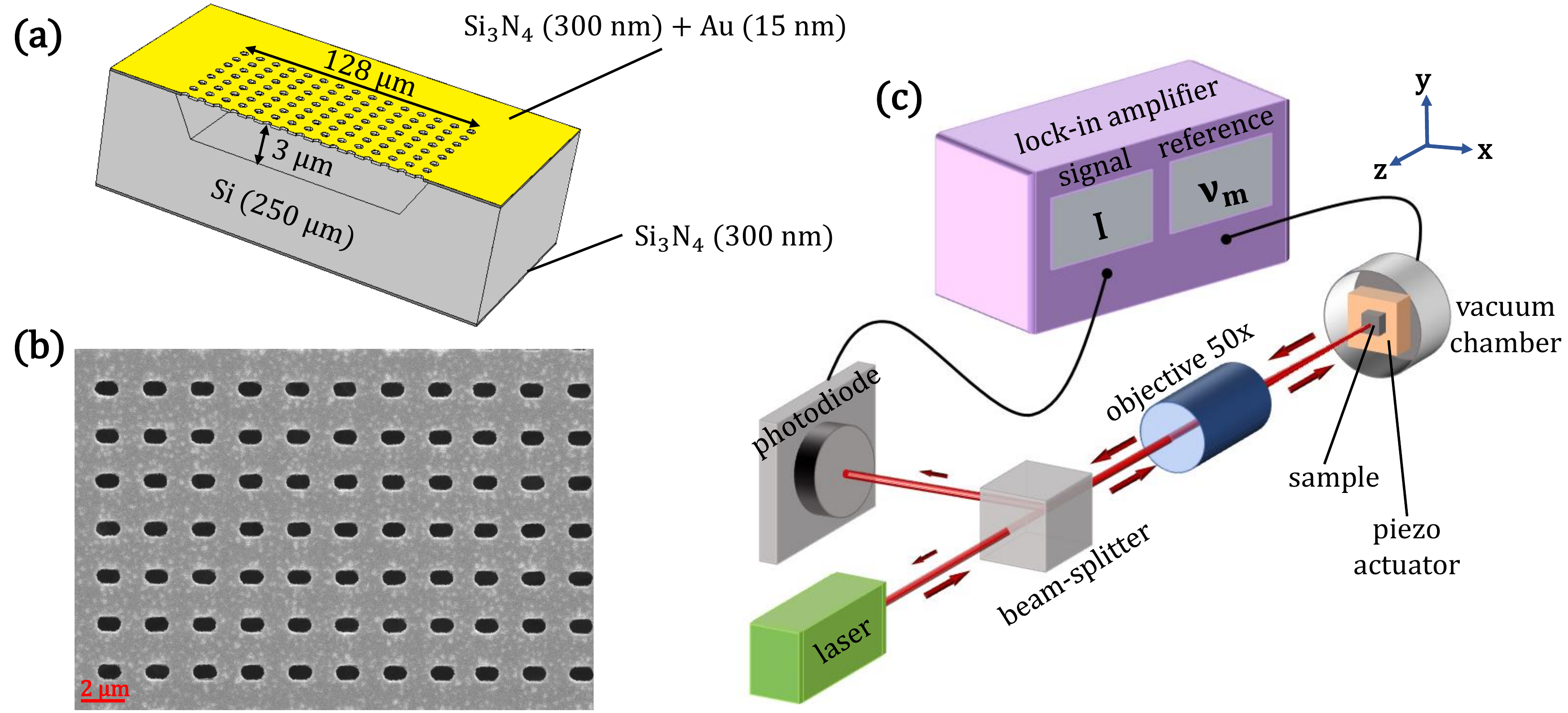}
	\caption{(a) Sketch of the cross section of a membrane device. (b) SEM image of a portion of the $\alpha=0.8$ membrane from the top. (c) Experimental setup employed for measurements.}
	\label{fig:2}
\end{figure*}

The metasurface consists of a 300-nm-thick Si$_3$N$_4$ slab perforated with holes, arranged in a square pattern with lattice constant $a = 2$ $\mu$m (see Fig. \ref{fig:1} (a)). The hole was designed starting from a rectangle with aspect ratio dependent on an asymmetry parameter $\alpha$; its sides were given by  $l_x=a(2-\alpha)/2$ and $l_y=a(\alpha)/2$, respectively. In order to avoid large stress accumulation at sharp corners, the hole was filleted using two circles of radius $l_y/2$ placed at the rectangle outermost edges in $\hat{x}$ direction. We investigated six different metasurfaces with $\alpha$ varying from 1 (perfectly circular holes) to 0.5; the unit cell of each membrane has been reported in Fig. \ref{fig:1} (b). The calculation of Block modes dispersion for an infinite metasurface was performed using a commercial Finite-Element Method (FEM) solver (COMSOL Multiphysics). Si$_3$N$_4$ was described using the material parameters reported in Table \ref{tab:1} (more on this later). The eigenfrequencies were determined in different points of the reciprocal space around $\Gamma$ which were selected using appropriate Floquet-Bloch boundary conditions. As expected in quasi-2D systems, the first simulated acoustic mode showed a quadratic dispersion, while the second and third had a linear one. The group sound velocities around a frequency of 2 MHz for different propagation directions are reported in the polar plots of panels (c-e) for the out-of-plane (ZA), transversal (TA) and longitudinal (LA) acoustic modes, respectively. The polar plot angle is relative to the $\hat{x}$ direction. Running simulations for decreasing $\alpha$, all the velocity plots became more anisotropic. The particular deformation of the LA mode offers an intuitive interpretation of the pattern effect, which results in a small elastic stiffening for the dispersion along $\hat{x}$, concurring to a strong softening in the one along $\hat{y}$ (see the arrows as guides for the eyes). This can be interpreted considering the elongated shape of the holes, which are more easily "squeezed" across the long edge rather than across the short one. Overall, these velocity plots suggest that the metasurface can be described by an artificial material with different sound velocities along the two directions. A metric for the material degree of anisotropy can be obtained through the Ledbetter-Migliori index, which is defined by the ratio of the square of the maximum and minimum shear velocities (TA) \cite{Ledbetter-Migliori}. Using the results reported in Fig. \ref{fig:1} (d), we obtained an anisotropy of 1.568 at $\alpha=0.5$. Starting from these results on infinite metasurfaces, we fabricated finite-size mechanical membranes with embedded patterns with varying $\alpha$s in order to experimentally study the effect of the artificial material on the device resonances.

The starting point for fabrication was a 250-$\mu$m-thick, double side polished, silicon wafer with a 300-nm-thick LPCVD high-stress Si$_3$N$_4$ film deposited on top and bottom surfaces. Using an AR-P 6200 resist mask, several periodic patterns with different asymmetry parameter $\alpha$ were defined via electron beam lithography on one side of the wafer, each one containing an array of 60x60 holes. The exposed Si$_3$N$_4$ was removed through reactive ion etching. Afterwards, the wafer was immersed in a hot (80$^\circ$C) 30\% in  weight KOH solution, so that the membranes were completely suspended upon etching of the Si underneath. The final membrane side was of 128 $\mu$m.  Finally, 15 nm of gold were thermally evaporated on the membranes in order to enhance their optical reflectivity. Realization of membranes with $\alpha<0.5$ was discarded after several attempts, as it resulted in a strong decrease of the fabrication yield due to the pattern-induced high stress on the membrane upon release. A cross sectional sketch of a membrane device is reported in Fig. \ref{fig:2} (a), while a scanning electron microscopy (SEM) of a detail of one fabricated membrane ($\alpha=0.8$) is reported in Fig. \ref{fig:2} (b).

\begin{figure*}[t]
	\includegraphics[width=16cm]{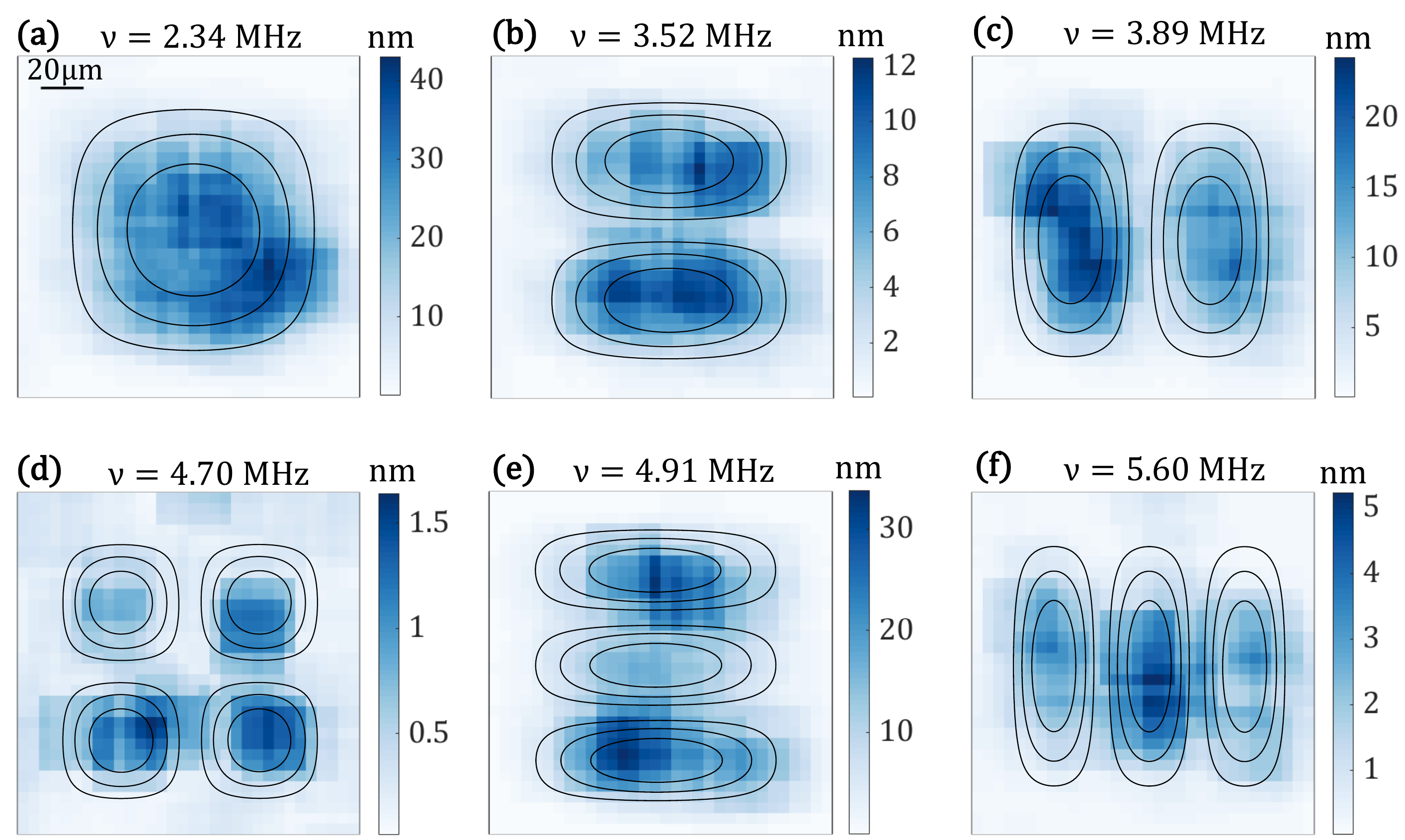}
	\caption{Maps of the mechanical modes of the membrane with $\alpha$=0.6. (a) Fundamental mode. (b)-(c) First excited modes. (d) Second excited mode. (e)-(f) Third excited modes. The colorscale depicts the out-of-plane displacement. The superimposed level curves were obtained through simulations.}
	\label{fig:3}
\end{figure*}

\begin{figure}
	\includegraphics[width=8cm]{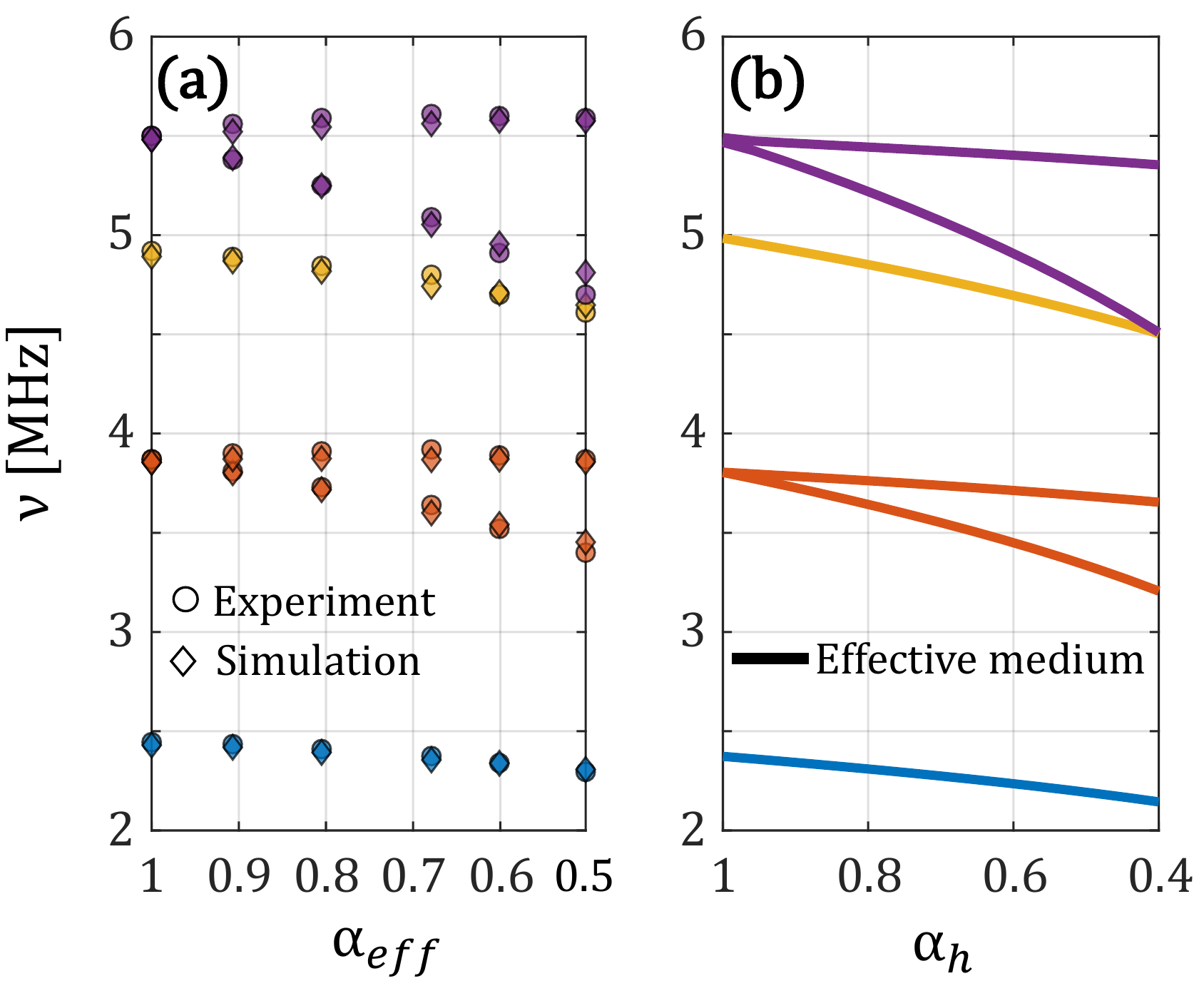}
	\caption{(a) Experimentally determined (circles) and simulated (diamonds) eigenfrequencies for each membrane. (b) Theoretically calculated frequencies of membranes composed of an effective medium with anisotropic material properties.}
	\label{fig:4}
\end{figure}

The characterization of the membrane mechanical modes was realized through self-mixing interferometry\cite{self-mixing}, a technique widely employed for vibration and displacement measurements of optomechanical devices\cite{Mezzapesa, blood, Huang, Baldacci, Ottomaniello, Vicarelli}. As shown in Fig. \ref{fig:2} (c), a $\lambda$ = 945 nm laser beam was focused upon the membrane through a 50x microscope objective, providing a roughly 36 $\mu$m beam spot size in the focal plane\cite{Vicarelli}. The laser signal was then reflected back into the laser cavity, where it interfered with itself. The reflected laser field carried information about membrane reflectivity and position: a careful treatment of the Lang-Kobayashi equations governing the system\cite{Baldacci, Vicarelli} can be used to show that membrane fluctuations in $\hat{z}$ correspond to light intensity modulations, which can be detected using an external photodiode. In order to selectively excite the mechanical modes, the sample was mounted on a piezoelectric actuator; this was driven with the signal generator of a lock-in amplifier, which was then used to demodulate the photodiode signal for coherent detection of membrane vibrations. By sweeping the frequency of the sinusoidal drive, resonance peaks appeared in the lock-in amplitude $R$, corresponding to the normal modes of oscillation of the membrane. The sample was kept in vacuum in order to reduce thin-film damping effects; the vacuum chamber was mounted on motorized translational $\hat{x}-\hat{y}$ stages, which were used for acquiring displacement maps. As an example, maps of mechanical modes of the $\alpha=0.6$ membrane are shown in Fig. \ref{fig:3}. Aided by displacement isolines obtained from simulations, we can clearly recognize the out-of-plane, lowest order mechanical modes, starting from the fundamental (a) to the first (b-c), second (d) and third excited mode (e-f). A rough checking of the measured frequencies with the velocities reported in Fig. \ref{fig:1} can be done by using the analytical formula for the angular frequency $\omega$ of the fundamental flexural mode of a square membrane\cite{Mech_Slater}:
\begin{equation}
\omega=\frac{2\pi v}{L}
\label{eq:mode}
\end{equation}
where $L$ is the membrane size and $v$ the sound velocity. Using the results of Fig. \ref{fig:3}, we found a velocity of about 300 m/s, which well compares with the velocity at 2 MHz of the ZA mode of Fig. \ref{fig:1} (c), also considering that we disregarded pre-stress effects in this comparison. Note that the lowest membrane modes originate from the ZA mode in the infinite metasurface, as can be intuited from the strong out-of-plane nature of the formers. The maps were deliberately oversampled, using a 5 $\mu$m pixel size, almost 7 times smaller than the laser spot size, in order to smooth the noise fluctuation generated by the self-mixing detection scheme. The diverse magnitude of the absolute value of out-of-plane displacements (obtained through a calibrated vibrometer, see Supplementary Material) suggests a different degree of coupling of the modes with the piezoelectric actuator, which did not have a white excitation spectrum as well as possibly being ill-coupled with modes with certain symmetries. Some of the modes, namely the first and third excited, are energy-split, suggesting a breaking of some system symmetry. By looking at the shape of the metasurface hole, we expect that for $\alpha<1$ the mirror symmetry across the square diagonal is broken: this is consistent with the fact that only first and third excited modes are split, while the second excited mode keeps its degeneracy. Similar maps were measured for all the six membranes. The eigenfrequencies are plotted in Fig. \ref{fig:4} (a) as a function of the effective asymmetry parameter $\alpha_{eff}$. This parameter was obtained from inspecting the SEM pictures of each membrane: more details have been reported in the Supplementary Material.

The experimentally observed membrane mechanical modes were simulated employing the FEM solver (COMSOL Multiphysics). The whole membrane, including the patterned holes, was reproduced in the simulations using a full three-dimensional domain with fixed boundary conditions. The actual sizes of the patterned features, corresponding to effective asymmetry parameter $\alpha_{eff}$, were employed to have a precise match with the experimental membrane. The material properties were embedded in Hooke's law for linear, elastic materials:
\begin{equation}
\varepsilon=S:\sigma+\sigma^{ex},
\label{eq:hooke}
\end{equation}    
where $\sigma$ and $\varepsilon$ are the stress and strain tensor, respectively. $\sigma^{ex}$ is an extra contribution possibly given by initial strain, as in the case of strongly pre-stressed materials such as Si$_3$N$_4$\cite{spiderweb, SiN-Painter, SiN_growth}. The constituent materials of the membrane, namely  Si$_3$N$_4$ and Au, are fully isotropic and can be described by the Young's modulus $E$ and Poisson's ratio $\nu$, which are embedded inside the compliance tensor $S$. The material density $\rho$ completes the set of parameters needed for the simulation. Table \ref{tab:1} reports the simulation parameters that best reproduced the experimentally observed modal frequencies.
\begin{table}[!htb]
	\begin{ruledtabular}
		\begin{tabular}{lcc}									             	& Si$_3$N$_4$ 	&	Au	\\
			\hline
			$\rho$ [kg/m$^3$] 	& 2370  & 19300 \\
			$E$ [GPa]			& 300	& 70 \\
			$\nu$ 				& 0.23  & 0.44 \\
			$\sigma^{ex}$ [GPa]	& 1.02	& 0 \\
			thickness [nm]      & 300   & 15 \\ 
		\end{tabular}
	\end{ruledtabular}
\caption{\label{tab:1} Simulation parameters.}
\end{table}
The simulations reproduced extremely well the experimental curves of the devices under investigation (semi-transparent diamonds of Fig. \ref{fig:4} (a)), confirming that the mode splitting was induced by and it is proportional to the pattern asymmetry.

In the metasurface approximation, the subwavelength membrane pattern mocks an effective medium, which in our case is an orthotropic mechanical material. This assumption can be validated by performing mechanical simulations of a homogeneous square membrane with the same lateral size as the investigated devices. The pattern effect can be introduced as a modification of the compliance $S$ and initial stress $\sigma^{ex}$.  
In an isotropic material, the 6x6 compliance tensor $S$ depends only on a set of two parameters; reduced symmetries require more independent parameters. For example, orthotropic materials, having symmetric properties about two or three mutually perpendicular planes, can be described by 9 independent parameters, namely the three components of $E$, $\nu$ and the shear modulus $G$, respectively. $S$ can be written in a compact matrix form using engineering notation:
\begin{equation}
S = 
\begin{pmatrix}
1/E_x							& -\nu_{yx}/E_y		& -\nu_{zx}/E_z	&	0	& 	0	&	0	\\
-\nu_{xy}/E_x	&	1/E_y							& -\nu_{zy}/E_z	&	0	& 	0	&	0	\\
-\nu_{xz}/E_x		&	-\nu_{yz}/E_y	& 1/E_z						&	0	& 	0	&	0	\\
0										& 	0										&	0									& 1/G_{xy}&	0						&	0	\\
0										& 	0										&	0									& 0						&	1/G_{yz}	&	0	\\
0										& 	0										&	0									& 0						&	0						&	1/G_{xz}	\\
\end{pmatrix}
\label{eq:ela_ortho}
\end{equation} 
In our specific case we assumed that the nanostructuration mainly affects the properties of one axis (i.e. $\hat{y}$), as can be hinted from the results of Fig. \ref{fig:1} (c). As a first approximation, we then considered the other two axes ($\hat{x}$ and $\hat{z}$) unaffected by the pattern shape. The anisotropy in the effective medium can be introduced by a parameter $\alpha_h$, which was used in an empirical procedure to rescale the Young's modulus in Eq. (\ref{eq:ela_ortho}). Furthermore, even if we considered full 3D simulations, we can assume that our devices are well within the plate approximation, meaning that they can be described using a reduced compliance matrix $S'$. Assuming that shear effects along the membrane thickness can be neglected (Kirchhoff-Love approximation \cite{Reddy}), $S'$ is composed only by $\hat{x}-\hat{y}$ planar components:
\begin{equation}
S' = 
\begin{pmatrix}
1/E_0																		&	-\nu_0/(\sqrt{\alpha_h}	E_0)		&	0	\\
-\nu_0/(\sqrt{\alpha_h}	E_0)		&	1/(\sqrt{\alpha_h} E_0)						&	0	\\
0																					&	0																	  			&	1/G_0	\\
\end{pmatrix}
\label{eq:ela_eff}
\end{equation}    
The last ingredient needed for the simulation of the effective medium is the inclusion of $\sqrt{\alpha_h}$ in the $\hat{y}$-component of the initial stress $\sigma^{ex}$. The simulated eigenfrequencies of a single 300-nm-thick layer as a function of $\alpha_h$ are reported in Fig. \ref{fig:4} (b). The use of the following parameters: $E_0$=300 GPa, $\nu_0$=0.23, $G_0$=122 GPa and $\sigma^{ex}$=500 MPa well reproduces the experimental spectroscopy of the metamaterial membrane; the comparison makes our patterned membrane fully equivalent to a homogeneous orthotropic layer with a tunable degree of anisotropy along $\hat{x}$ and $\hat{y}$, respectively.

\begin{figure}[h!]
	\includegraphics[width=6cm]{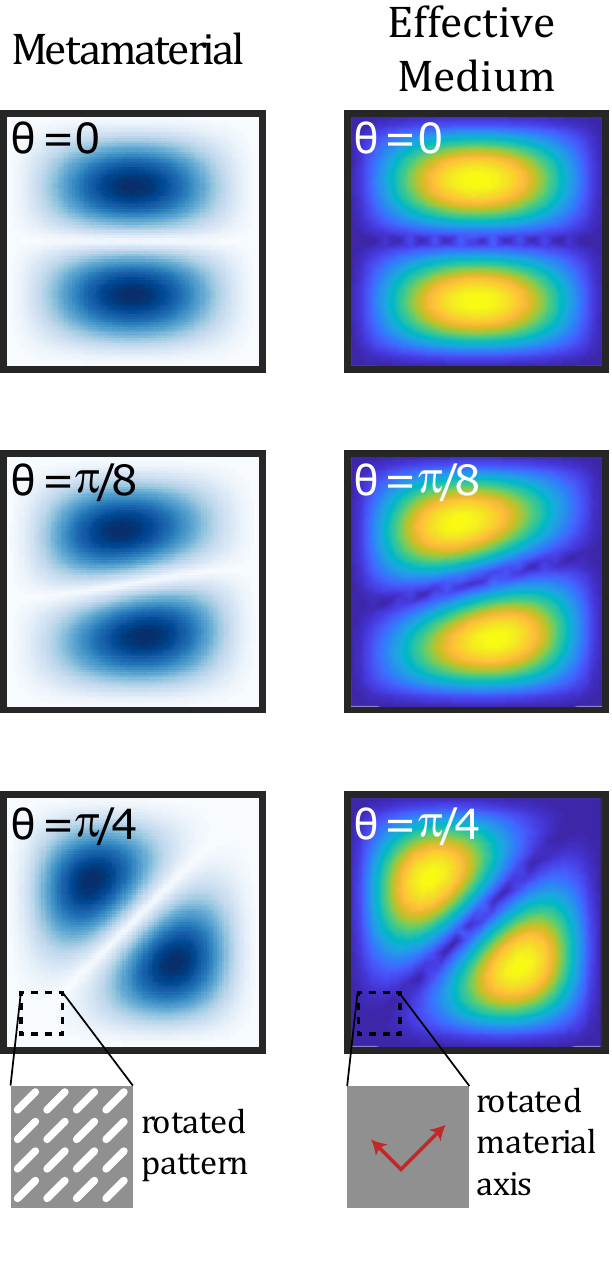}
	\caption{Comparison between the simulated first excited mode in a metamaterial membrane with rotated pattern with $\alpha=0.5$ (left panel) and an effective medium with rotated anisotropy axis (right panel).}
	\label{fig:5}
\end{figure}

The power of our approach relies in the possibility to design arbitrary, in-plane asymmetry axes, without specific restrictions to the directions parallel to the membrane sides, as in the case investigated so far. This favorably compares with other mechanical systems where symmetry breaking has been achieved by boundary modification (i.e. going from square to rectangular membranes\cite{carbon_modes}) or by the use of natural materials whose choice of orientation is often limited by fabrication constraints (i.e. material growth, anisotropic wet etching, etc.). A simple rotation of the geometrical pattern shown in Fig. \ref{fig:1} induces a rotation of the orthotropic axis, which can now be oriented in an arbitrary direction on the membrane $\hat{x}-\hat{y}$ plane. FEM simulations of the first excited mode for a model membrane with rotated pattern are reported in Fig. \ref{fig:5} for selected angles of rotation $\theta$. Here the asymmetry parameter has been taken equal to 0.5. As one can see, the mode shape follows closely the rotation angle, going from a mode with $\hat{y}$-mirror symmetry ($\theta$=0°) to a mode with mirror symmetry along the square diagonal ($\theta$=45°). Interestingly, this corresponds to an effective medium where both the elasticity matrix and initial stress have been rotated by the same quantity $\theta$. The simulation results for this latter case have been reported in the right panel of Fig. \ref{fig:5}.        

In this Letter we have introduced artificial degrees of anisotropy by micro-patterning a homogeneous Si$_3$N$_4$ layer. By controlling the pattern shape asymmetry, we can modify the sound velocity along a specific crystallographic orientation. Holding a maximum Ledbetter-Migliori anisotropy of 1.568, our artificial layer can be used to control the symmetry-breaking mode splitting in square membranes, allowing tuning of the resonances to address particular device parameters. Additionally, the artificial anisotropy can be realized in arbitrary directions on the membrane plane, with a further tuning knob for membrane design. Our approach flexibility makes it useful for the implementation of multi-mode devices for mass, force and acceleration sensing as well as for the realization of optimized systems for opto- and electro-mechanics, where mechanical resonators can interact with electromagnetic waves, for example in membrane-in-the-middle configurations. Here, mode control can be useful for enhancing or suppressing the interaction of mechanical modes with electromagnetic cavity modes of certain symmetries. 

\section*{Supplementary Material}
See Supplementary Material for the calculation of the effective asymmetry parameter $\alpha_{eff}$ and for the estimation of the out-of-plane displacements.
		
\section*{Author declarations}
\subsection*{Conflict of interest} 
The authors declare no conflict of interest.

\section*{Data Availability Statement}
The data supporting the findings of this study are available from the corresponding author upon reasonable request.

	\end{document}